\documentclass[twocolumn,showpacs,preprintnumbers,amsmath,amssymb]{revtex4}

\usepackage{graphicx}
\usepackage{dcolumn}
\usepackage{bm}
\usepackage{amsmath}
\usepackage{times}
\usepackage[usenames]{color}
\begin{document}

\title{Ultrafast Spectral Weight Transfer in $R$BaCo$_2$O$_{6-\delta}$ ($R$=Sm, Gd, and Tb): A Role of Electronic Correlation in Photoinduced Phase Transition}
\author{Y. Okimoto${\rm ^{1}}$, T. Miyata${\rm ^{1}}$, M.S. Endo${\rm ^{1}}$, M. Kurashima${\rm ^{1}}$, K. Onda${\rm ^{2}}$, T. Ishikawa${\rm ^{1}}$,
S. Koshihara${\rm ^{1,3}}$, M. Lorenc${\rm ^{4}}$, E. Collet${\rm ^{4}}$, H. Cailleau${\rm ^{4}}$, and T. Arima${\rm ^{5}}$} 
\affiliation{$^1$Department of Materials Science, Tokyo Institute of Technology, Meguro, Tokyo, 152-8551, Japan}
\affiliation{$^2$Department of Environ. Chem. and Engineering, Tokyo Institute of Technology, Nagatsuta, Yokohama, 226-8503, Japan}
\affiliation{$^3$CREST, JST, Chiyoda-ku, Tokyo 102-0075, Japan}
\affiliation{$^4$Institut de Physique de Rennes, UMR 6251, CNRS-Universite, 35042, Rennes Cedex, France}
\affiliation{$^5$Institute of Multidisciplinary Research for Advanced Materials, 
Tohoku University, Sendai, 980-8577, Japan}

\date{\today}
\begin{abstract}
We performed femtosecond reflection spectroscopy on a series of perovskite-type cobalt oxide $R$BaCo$_2$O$_{6-\delta}$ ($R$=Sm, Gd, and Tb) crystals, in which the electronic transfer was controlled by $R$.
 The transient reflectivity and the optical conductivity ($\sigma^{\rm PI}(\omega)$) obtained by Kramers-Kronig analysis showed an ultrafast change within a time resolution ($\approx 150$ fs) at room temperature and the appearance of signals of a hidden state different from the high temperature metallic state. 
The transferred spectral weight in $\sigma^{\rm PI}(\omega)$ upon photoexcitation sensitively depended on the $R$-species, indicating an important role of electronic correlation in the photoexcited state.
\end{abstract}

\maketitle

%
\narrowtext

The photonically excited states in condensed matter have been extensively studied, stimulated by recent developments in laser technology. 
One of the most important and intriguing motivations in such studies is to actively create and control a new phase concealed in the material by the irradiation of laser light. 
This is termed photoinduced phase transition (PIPT) and has recently been noticed as an important research topic in nonequilibrium physics\cite{Nasu}. 
Several interesting examples using various optical techniques\cite{Cavalleri} have been demonstrated, especially in strongly correlated materials, which are subjective to external stimuli due to the close interaction between many degrees of freedom such as charge, spin, and lattice.
 In addition to experimental findings, theoretical approaches for elucidating the PIPT phenomena have been reported by exactly resolving the effective Hamiltonian in each target system \cite{theory}.

\begin{figure}[t]
\includegraphics[width=\columnwidth,clip]{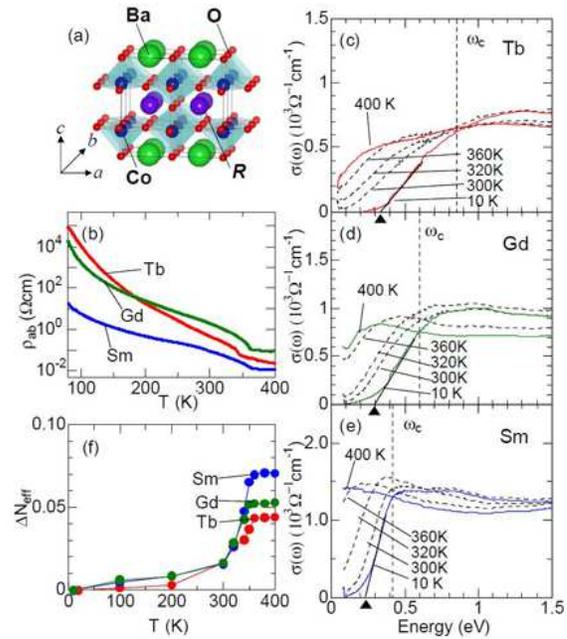}
\caption{(color online) (a): Schematic drawing of the crystal structure in GdBaCo$_2$O$_{6-\delta}$ after [7].
(b): Temperature dependence of resistivity ($\rho$) in $R$BaCo$_2$O$_{6-\delta}$ ($R$=Sm, Gd, and Tb). 
(c)-(e): Temperature dependence of optical conductivity ($\sigma(\omega)$) in $R$BaCo$_2$O$_{6-\delta}$ for $R$=Tb (c), Gd (d), and Sm (e).
(f): Temperature dependence of Drude weight, $D$ calculated from $\sigma(\omega)$ in $R$BaCo$_2$O$_{6-\delta}$.}
\label{fig:fig1}
\end{figure}

Among these approaches, spin-crossover (SC) Fe complexes are known as typical systems for the PIPT phenomena\cite{SC}. This is a reversible photocontrol of the magnetic state of $d$ electrons in the strong ligand field between two states; the low spin (LS) state ($t_{2g}^6$) stabilized by the crystal field and the high spin (HS) state ($t_{2g}^4 e_g^2$) by Hund's coupling. 
In SC materials, the energy difference between the LS and HS states is so subtle that photoirradiation as well as temperature change can switch the spin configuration.

Another noted SC system is a class of perovskite-type cobalt oxide, and one such example is Pr$_{0.5}$Ca$_{0.5}$CoO$_{3}$, which undergoes a phase transition between the insulating LS state and the itinerant intermediate spin (IS) state ($t_{2g}^5 e_g^1$) in the Co$^{3+}$ site \cite{Tsubouchi}. 
In this material, a photonic phase control in femtoseconds has recently been reported\cite{Okimoto1}, which is an ultrafast insulator-metal (I-M) transition involved in the spin state transition between the LS and the IS state.
This is in contrast to the simple magnetic PIPT observed in Fe complexes\cite{SC}. 
This implies that the perovskite cobalt oxide systems containing trivalent Co ions offer us a unique arena for studying PIPT as a strongly correlated SC system.

From this point of view, comparison with other cobaltite compounds, $R$BaCo$_2$O$_{6-\delta}$ ($R$BCO, $R$=Sm, Gd, and Tb) can give important insight.
 The crystal structure of $R$BCO is shown in the inset of Fig. 1(a)\cite{Moritomo}. 
The $R$-O and Ba-O layers are alternately stacked along the $c$-axis, forming a quasi-2D structure.
After annealing without any special condition, the $\delta$ value denoting oxygen deficiency was estimated as about $\approx 0.5$.
 Hence, Co ions are almost trivalent (Co$^{3+}$).
As seen in the schematics, there are two different Co sites, octahedral and pyramidal sites. 
While the spin state of Co$^{3+}$ in the pyramidal site is always in the IS state with the configuration of ($t_{2g}^5 e_g^1$) due to the anisotropic ligand field, a thermally induced spin state change from the LS to HS states occurs in the octahedral sites above room temperature\cite{Frontera}.

 In Fig. 1(b), we show temperature dependence of resistivity ($\rho$) in $R$=Tb, Gd, and Sm.
 In the three materials, the $\rho-T$ curves show a sudden jump at around $T_{\rm IM}\approx 350$ K, indicating an I-M transition \cite{Takubo}.
Upon the I-M transition, volume expansion occurs at $T_{\rm IM}$, driven by spin excitation from the LS ground state\cite{Frontera}, a characteristic feature of the SC system.

We performed time-resolved reflection spectroscopy on a series of $R$BCO ($R$=Sm, Gd, and Tb) crystals using a femtosecond laser system and investigated the ultrafast dynamics in the photo-excited state at room temperature.
 There were two purposes of this study; one was to understand the electronic structure in the photoexcited as well as the thermally induced metallic states in the cobalt oxide, and the other was to systematically reveal the effect of electronic correlation on the appearance of the photoexcited state by changing the $R$ site in $R$BCO crystals, in terms of experimental point of view. 
 
Single crystals of $R$BCO were grown using a floating zone method\cite{Saito}.
 The reflectivity spectrum ($R(\omega)$) for the light polarization perpendicular to the $c$-axis was measured using a Fourier transform-type interferometer (0.01-0.7 eV) and grating monochromator (0.6-5 eV). 
The relative change in reflectivity ($\Delta R/R$) after photoirradiation was obtained with the conventional pump-probe method using a Ti:sapphire regenerative amplifier system (pulse width $\approx$150 fs, repetition rate 1 kHz, and photon energy 1.58 eV) as a light source.
 The amplified light was separated into two light beams.
 We used one beam as a pump light for excitation. 
The photon energy of a pump light (1.58 eV) corresponds to the excitation energy from the O $2p$ to Co $e_{g}$ bands\cite{Saito}.
 The other beam, whose frequency was converted with an optical parametric amplifier was used as a probe light to investigate transient $R(\omega)$ after photoexcitation from 0.12 to 2.1 eV . 
\begin{table}[tbp]

\label{tbl_list}

\caption{
Physical parameters for $R$BaCo$_2$O$_{6-\delta}$.
}
\begin{center}
\begin{tabular}{lrrr}
\hline
$R^{3+}$  & Sm & Gd & Tb \\
\hline
Ionic radius (\AA) & 1.240 & 1.215 & 1.203\\
Cut off energy (eV) & 0.36 & 0.52 & 0.80 \\
Penetration depth (\AA) & 470 & 600 & 700 \\
Optical gapieVj & 0.23 & 0.30 & 0.34 

\end{tabular}
\end{center}

\end{table}

First and foremost, we discuss the role of $R$ in the electronic structure. In the three dimensional perovskite-type structure, the reduction of the ionic size of $R$ generally brings about the lattice distortion and the resultant suppression of electron transfer ($t$)\cite{Torrance}, and this can also be expected in layered $R$BCO structure.
 Such an effect of the $R$-size on the electronic structure in $R$BCO was confirmed using optical spectroscopy. 
Figures 1(c)-(e) show temperature dependence of the optical conductivity spectra ($\sigma(\omega)$s) in $R$=Sm, Gd, and Tb crystals. At 10 K, all the spectral shapes of $\sigma(\omega)$ are gap-like, reflecting the insulating ground state. We estimated the optical gap ($\Delta_g$) by linear extrapolation from the rising part of the $\sigma(\omega)$ to the abscissa, as shown by the solid lines and triangles.
 The evaluated values are listed in Table 1 with the ionic radii of $R^{3+}$ after ref\cite{Shannon}.
 The value of $\Delta_g$ becomes large with decreasing the ionic radius.

\begin{figure}[t]
\includegraphics[width=\columnwidth,clip]{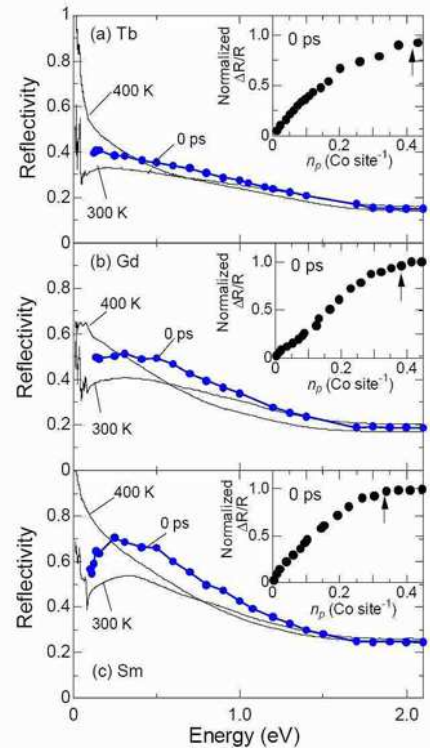}
\caption{(color online) (a)-(c): Transient reflectivity spectra just after the photoirradiation in $R$BaCo$_2$O$_{6-\delta}$ for $R$=Tb (a), Gd (b), and Sm (c) (black circles).
 The solid lines show linear reflectivity spectra at 300 K and 400 K (metallic state).
 The insets denote the fluence dependence of the relative change of reflectivity at 0.75 eV.}
\label{fig:fig2}
\end{figure}

With increasing temperature, in all the three samples, the onset energy of the absorption shows gradual redshift, and finally a Drude-like metallic spectrum appears at 400 K above $T_{\rm IM}$. 
To see the thermal spectral weight transfer more quantitatively, we have calculated the effective number of electrons ($N_{\rm{eff}}$), defined as

\begin{equation}
 N_{\rm{eff}} = \frac{2m_{0}}{\pi e^{2} N} \int_{0}^{\omega_{c}}\sigma(\omega)d\omega,
\end{equation}
where $m_0$ and $e$ are mass and charge of an electron, respectively, and $N$ is the number of Co ions in unit volume. For estimating the contribution from the inner gap excitation (i.e., Drude weight), we defined the cut off energy ($\omega_c$) as the crossing point in $\sigma(\omega)$ spectra at 10 K and 400K, as shown in Figs. 1(c)-(e), and calculated $D$, the increased value in $N_{\rm{eff}}$ from the ground state with temperature (i.e., $D =N_{\rm{eff}}$($T$)- $N_{\rm{eff}}$($T =10$ K)). Figure 1(f) plots the temperature dependence of $D$. The value of $D$ suddenly increases above $T_{\rm IM}$, reflecting the I-M transition driven by the thermally excited $e_g$ carriers. 
An important feature is that $D$ in the metallic state strongly depends on the species of $R$.
 With decreasing the ionic size of $R$ from $R$=Sm to Tb, $D$ decreases, indicating the suppression of the Drude weight due to the $e_g$ carriers.
The decrease in the ionic size brings about suppression of $t$ for $e_g$ electrons ($t_{\rm A}$) through the lattice distortion, and hence it is reasonable to consider that $D$ decreases with decrease in the value of $t_{\rm A}$.
Hereafter, we demonstrate results of femtosecond spectroscopy in the cobalt system and discuss $t_{\rm A}$ dependence in the ultrafast dynamics by changing the species of $R$.

In the insets of Figs. 2(a)-(c), we show the fluence dependence of $\Delta R/R$ at $\approx$0 ps in $R$=Sm, Gd, and Tb at 0.75 eV. 
In all the samples, $\Delta R/R$ shows a gradual increase with increasing fluence. Such a linear increase of $\Delta R/R$ without threshold is similar to the case of manganese oxides\cite{Okimoto2}. 
With further increase in the fluence, $\Delta R/R$ shows a saturation at around the excitation intensity ($I_0$) denoted by an arrow in each of these figures. 
This indicates that the photon number necessary for the complete photoinduced change becomes larger with decreasing $t_{\rm A}$.


 Figure 2 shows the transient $R(\omega)$ spectrum in each sample ($\approx$0 ps) obtained with the fluence of $I_0$, the maximum variation just after photoirradiation. For reference, we also plot $R(\omega)$ in the equilibrium states at room temperature and 400 K, denoted by solid lines (the spiky structures below 0.1 eV are due to optical phonon modes).
 As the temperature increases, $R(\omega)$ at lower energies increases  and a Drude-like spectrum appears in the metallic phase (400 K). As can be clearly seen, the spectral shapes of the transient $R(\omega)$s at $\approx$0 ps are completely different from those at 400 K. 
This clearly indicates that simple heating by laser irradiation cannot explain the reflectance change. Another important aspect is the degree of change in the transient $R(\omega)$ ($\Delta R$) after photoirradiation. The $\Delta R$ value increases with an increase in $t_{\rm A}$. 
To see the $t_A$-dependence in the optical spectra more clearly, we calculated transient conductivity at $\approx$0 ps ($\sigma^{\rm PI}(\omega)$) by means of Kramers-Kronig (K-K) analysis considering the spatial variation of a dielectric function along the depth of the sample\cite{cal}.

 In Figs. 3(a)-(c), we show the calculated $\sigma^{\rm PI}(\omega)$ in $R$=Sm, Gd, and Tb at $\approx$0 ps (black circles), together with linear $\sigma(\omega)$ at 300 and 400 K. In all the materials at $\approx$0 ps, the spectral weight of $\sigma^{\rm PI}(\omega)$ in the mid-infrared region increases, forming a broad peak. Such a transient spectral shape is completely different from that in the metallic state, which strongly indicates that photoexcitation created a new phase that cannot be observed by thermally induced I-M in accordance with the spin state transition.


In SC Fe complexes, photoexcitation causes the spin state transition between the LS and HS states and several theoretical frameworks have been reported\cite{SC}. 
However, in the present SC Co system, which shows a different phase from the high temperature state, a different theoretical viewpoint regarding the strong correlation between $e_g$ carriers is indispensible. Recently, Kanamori, Matsueda, and Ishihara have proposed an effective Hamiltonian to describe the spin state transition in a SC cobalt system using a two-band Hubbard model\cite{Kanamori}. 
They have numerically calculated  the ground state after injecting a pair of electron and hole in the effective Hamiltonian and revealed that the HS-hole bound state as a result of the photoexcitation was stabilized which shows a nonmetallic absorption peak in $\sigma^{\rm PI}(\omega)$, with the Drude component due to the free electrons and holes in the lower energy region. 
Such a bound state, which is stabilized by the local double exchange interaction, seems to be consistent with the observed photoinduced state having a clear mid-infrared peak different from the Drude absorption.

\begin{figure}[t]
\includegraphics[width=\columnwidth,clip]{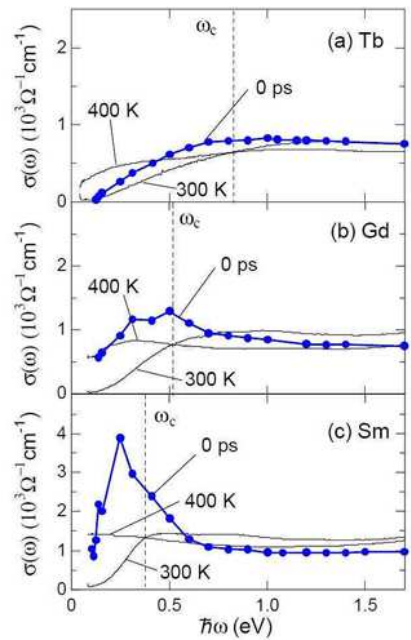}
\caption{(color online) (a)-(c): Transient optical conductivity spectra ($\sigma^{\rm PI}(\omega)$) just after the photoirradiation in $R$BaCo$_2$O$_{6-\delta}$ for $R$=Tb (a), Gd (b), and Sm (c) (black circles). The solid lines show linear$\sigma(\omega)$ without photoirradiation at 300 K and 400 K.}
\label{fig:fig3}
\end{figure}

Finally, we discuss the increased spectra weight at room temperature just after photoirradiation ($\Delta N_{\rm{eff}}$), defined as 


\begin{equation}
 \Delta N_{\rm{eff}} = \frac{2m_{0}}{\pi e^{2} N} \int_{0}^{\omega_{c}}\sigma^{\rm PI}(\omega)d\omega -
 N_{\rm{eff}}(T=\rm{300 K}).
\end{equation}
We plotted $\Delta N_{\rm{eff}}$ and $D$ as a function of the ionic radius of $R^{3+}$ ion in Fig. 4. 
By photoexcitation, $\Delta N_{\rm{eff}}$ almost linearly increases with increasing the ionic size of $R^{3+}$, i.e., $t_{\rm A}$, which strongly indicates that electronic correlation seriously affects the appearance not only of the thermally induced state but also of the photoinduced state.
In addition, while the value of $\Delta N_{\rm{eff}}$ is comparable to $D$ in $R$=Tb, $\Delta N_{\rm{eff}}$ becomes larger than $D$ with further increasing $t_{\rm A}$, indicating the larger spectral weight transfer by the photoexcitaiton than by the thermal change.
 It is worth noting here that the above mentioned calculation\cite{Kanamori} also reproduces the two  experimentally observed features; i.e., the photonically induced oscillator strength in $\sigma(\omega)$, which originated from the HS-hole bound state, increases with $t_{\rm A}$, and is larger than the thermally induced spectral weight, supporting the assignment of the mid-infrared peak in $\sigma^{\rm PI}(\omega)$. 
 
\begin{figure}[t]
\includegraphics[width=\columnwidth,clip]{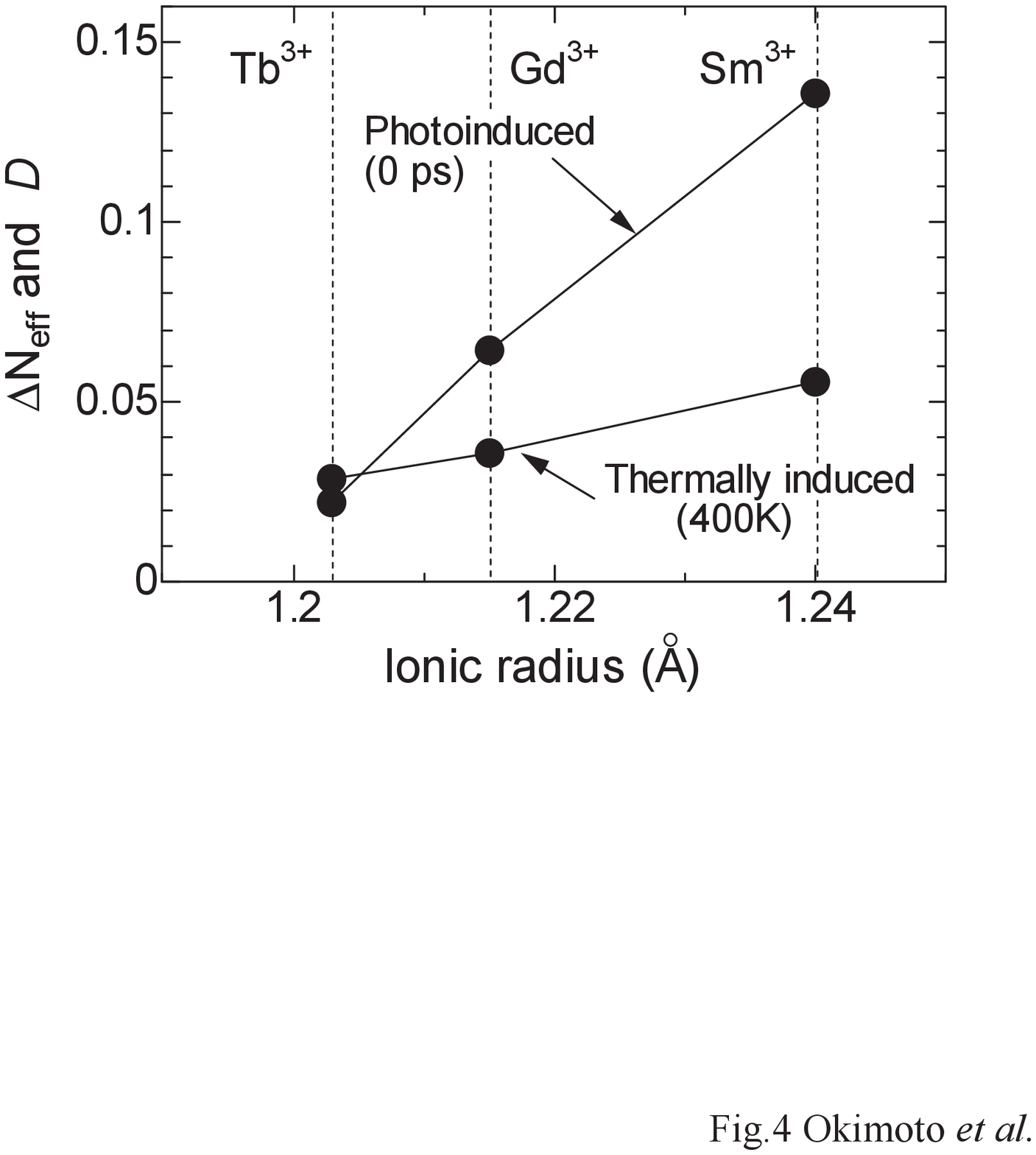}
\caption{The variation of the photoinduced ($\Delta N_{\rm{eff}}$) and thermally induced ($D$) spectral weight transfer as function of the ionic raidus of $R^{3+}$ in $R$BaCo$_2$O$_{6-\delta}$. }
\label{fig:fig4}
\end{figure}

In summary, we performed femtosecond reflection spectroscopy in perovskite-type cobalt oxide, where the electronic transfer $t_{\rm A}$ is controlled by changing the rare earth species. 
The transient $R(\omega)$ as well as the calculated $\sigma^{\rm PI}(\omega)$ showed an ultrafast change at room temperature, indicating (1) the appearance of a hidden state just after photoirradiation different from the high temperature metallic state, which is assigned as the HS-hole bound state suggested by the recent theoretical calculation, and (2) that the transferred spectral weight by photoexcitation increases with increasing $t_{\rm A}$, which is also consistent with theoretical prediction. These results indicate an important role of electronic correlation in photoinduced phase transition, which not only sheds important insight on understanding the photoexcited state but also provides a good stage exploring novel ultrafast nonequilibrium phenomena.

\par

The authors thank T. Saito for technical assistance and Y. Kanamori, H. Matsueda, and S. Ishihara for fruitful discussions. This work was supported by a Grant-in-Aid (Grants No. 21104514) for Scientific Research from the Ministry of Education, Culture, Sports, Science, and Technology, Japan.

\vfill
\eject

\begin{references}
\bibitem{cal}In the calculation of $\sigma^{\rm PI}(\omega)$, we assume that the photoinduced state occurs in proportion to the intensity of the pump light exponentially decayed along the depth direction of the sample\cite{Okimoto2}. The penetration depths are listed in Table 1.
\end{references}
\end{document}